\documentclass[12pt,twoside]{article}
\usepackage[mathscr]{eucal}
\usepackage{amsmath,amsfonts,amssymb,amsthm,mathabx,empheq}
\usepackage{times}
\usepackage{pdfsync}
\usepackage{cite}
\usepackage{url}
\usepackage{hyperref}
\usepackage{tensor}
\usepackage{color}
\usepackage{multicol}
\usepackage{bbold}
\usepackage[thicklines]{cancel}

\advance\textheight\topskip
\allowdisplaybreaks[1]

\newcommand\td{\text{d}}
\newcommand\cO{{\cal O}}

\newcommand{\p}{\partial}

\newcommand{\be}{\begin{equation}}
\newcommand{\ee}{\end{equation}}
\newcommand{\bea}{\begin{eqnarray}}
\newcommand{\eea}{\end{eqnarray}}

\def\bg{\bar g}

\def\blue{\textcolor{blue}}

\newcommand{\nn}{\nonumber}
\newcommand*\xbar[1]{%
  \hbox{%
    \vbox{%
      \hrule height 0.5pt 
      \kern0.3ex
      \hbox{%
        \kern-0.0em
        \ensuremath{#1}%
        \kern-0.0em
      }%
    }%
  }%
}

\allowdisplaybreaks[1]

\DeclareFontFamily{OT1}{rsfs}{} \DeclareFontShape{OT1}{rsfs}{m}{n}{
<-7> rsfs5 <7-10> rsfs7 <10-> rsfs10}{}
\DeclareMathAlphabet{\mycal}{OT1}{rsfs}{m}{n}

\hypersetup{
colorlinks=true,
linktoc=page,    
linkcolor=blue,
citecolor=blue,
urlcolor=blue,
colorlinks=true,
}

\begin{document}

\title{Perturbations of a Schwarzschild black hole and Newman-Unti gauge}

\author{Pujian Mao, Baijun Zeng, and Kai-Yu Zhang}
\date{}

\def\mytitle{Perturbations of a Schwarzschild black hole and Newman-Unti gauge}

\addtolength{\headsep}{4pt}

\begin{centering}

  \vspace{1cm}

  \textbf{\large{\mytitle}}

  \vspace{1cm}

  {\large Pujian Mao$^{\yen}$, Baijun Zeng$^{\yen}$, and Kai-Yu Zhang$^\text{\texteuro}$}

\vspace{0.5cm}

\begin{minipage}{.9\textwidth}\small \it  \begin{center}
     $^{\yen}$Center for Joint Quantum Studies, Department of Physics,\\
     School of Science, Tianjin University, 135 Yaguan Road, Tianjin 300350, China\\
     $^\text{\texteuro}$Institute for Theoretical Physics, TU Wien, \\
     Wiedner Hauptstrasse 8–10/136, A-1040 Vienna, Austria
 \end{center}
\end{minipage}

\end{centering}

\begin{center}
Emails:  pjmao@tju.edu.cn, zeng@tju.edu.cn, kaiyu.zhang@student.tuwien.ac.at
\end{center}

\begin{center}
\begin{minipage}{.9\textwidth}
\textsc{Abstract}: In this paper, we derive the complete transformations of a generic first order perturbative metric in Schwarzschild spacetime to the Newman-Unti (NU) gauge in series expansion near null infinity. This allows us to determine the asymptotic shear, the mass aspect and the angular momentum aspect of the perturbative fields. As a direct application, we derive the corresponding asymptotic NU data for quasinormal modes (QNMs) of a Schwarzschild black hole. The total energy and angular momentum of QNMs obtained from the asymptotic NU data vanish but classical supertranslation charges are non-trivial, which can be applied to fix the BMS frame at first perturbative order. 

\end{minipage}
\end{center}

\thispagestyle{empty}

\newpage
\tableofcontents

\section{Introduction}

Gravitational-wave physics has been always at the center of attention for many researchers since the first direct detection of gravitational waves by the LIGO and Virgo collaborations \cite{LIGOScientific:2016aoc}. Matched filtering plays a central role in the detection of gravitational-wave signals buried in Gaussian noise \cite{Creighton:2011zz}. The effectiveness of this technique relies on the availability of accurate gravitational-wave waveform templates. Such templates are typically constructed using perturbative methods that provide quantitative predictions for the emitted gravitational radiation. Prominent examples include the post-Newtonian formalism \cite{Blanchet:2013haa}, the post-Minkowskian (PM) expansion \cite{PM} and black-hole perturbation theory \cite{Chandrasekhar,Pani:2013pma,Pound:2021qin}.

Historically, the existence of gravitational waves was still debatable up until the 1960s, largely because most analyses relied on linearized approximations of general relativity. This issue was resolved in the seminal works of Bondi, van der Burg, Metzner, and Sachs \cite{Bondi:1962px,Sachs:1962wk}, who formulated the Einstein equations as a characteristic initial-value problem. Within their framework, known as Bondi-Sachs (BS) formalism \cite{Madler:2016xju}, gravitational radiation is characterized by the news function near null infinity and the mass of the gravitational system decreases whenever the news function is non-vanishing. Similar conclusions were obtained by Newman and Unti (NU) \cite{Newman:1962cia} within the Newman-Penrose formalism \cite{Newman:1961qr}. The relation between these two frameworks has been studied extensively, see, for example \cite{Barnich:2011ty}.

At the technical level, both the BS framework and the NU formulation are based on asymptotic analysis. Consequently, they provide a characterization of gravitational radiation at null infinity rather than a complete description of the source dynamics in the bulk of the spacetime. In particular, the waveform generated by a given source cannot be determined directly within the asymptotic framework alone. Nevertheless, the asymptotic framework provides well-defined notions of asymptotically conserved quantities, including the four-momentum and angular momentum of radiating gravitational systems \cite{Bondi:1962px,Sachs:1962wk,Newman:1965ik,Newman:1968uj,Winicour,Geroch:1977big,Prior,Ashtekar:1978zz,Ashtekar:1979xeo,Ashtekar:1979iaf,Ashtekar:1981bq,Geroch:1981ut,Dray:1984rfa}; see also the more recent developments in \cite{Barnich:2010eb,Barnich:2011mi,Flanagan:2015pxa}. These notions play a key role in constructing accurate gravitational-wave templates \cite{Blanchet:2013haa} and in resolving conceptual puzzles associated with radiated observables. For example, in the PM analysis of gravitational scattering \cite{Veneziano:2022zwh}, the leading contribution to angular-momentum loss may appear at either $\mathcal{O}(G^2)$ or $\mathcal{O}(G^3)$, depending on the asymptotic gauge choice at null infinity and the corresponding definition of angular momentum.

Explicit relations between perturbative methods and asymptotic frameworks provide a crucial bridge between the quantitative and qualitative descriptions of gravitational radiation. Recently, the explicit transformation between the PM expansion in harmonic gauge and the asymptotic expansions in NU gauge was derived in \cite{Blanchet:2020ngx,Blanchet:2023pce}, primarily addressing the multipolar PM approximation. Furthermore, the complete transformation from generalized harmonic coordinates to NU coordinates for a generic perturbative metric at second PM order was presented in \cite{Mao:2025lwk}. Explicit connections between asymptotic frameworks and black-hole perturbation theory have also attracted considerable attention. Earlier studies investigated Schwarzschild perturbations in the BS gauge \cite{Bishop:2004ug}, and the transformation of standard Schwarzschild perturbations into the BS frame was explicitly constructed for the $\ell=2$ sector \cite{Kubeka:2014}. Quasinormal modes (QNMs) were subsequently investigated within the linearized BS framework \cite{Bishop:2009ba,Mongwane:2024vao}. More recently, a systematic formulation of black-hole perturbation theory in BS gauge, together with a prescription for fixing the associated BMS frame on a Kerr background, was developed in \cite{Spiers:2026yqx}.

In this work, we derive the coordinate transformations that map generic first order metric perturbations of Schwarzschild spacetime into NU gauge. We assume that the coordinate transformations are themselves expanded perturbatively and that the first order perturbative metric admits an expansion in inverse powers of the radial coordinate at large distances. We explicitly obtain the coordinate transformations at first perturbative order and subsequently obtain the asymptotic form of the perturbative metric in NU gauge. In particular, we extract the asymptotic NU data, including the asymptotic shear, mass aspect, and angular momentum aspect.

The background Schwarzschild mass parameter $M$ gives rise to additional contributions to the asymptotic NU data. These $M$-dependent terms vanish in the Minkowski limit and therefore have no analogue in the Minkowski-background analysis of \cite{Mao:2025lwk}. We also note that, to the best of our knowledge, Ref. \cite{Spiers:2026yqx} did not explicitly express the final metric in BS gauge in terms of the original metric components. Moreover, while the BS and NU frameworks are closely related, they are not identical. Therefore, the results presented here in the NU framework cannot be regarded as a special case of that work.

As a direct application, we obtain the asymptotic NU data associated with the QNMs of a Schwarzschild black hole. The corresponding asymptotically conserved charges are investigated in detail. In particular, the total energy and angular momentum vanish at the linear order. The mass aspect in the odd-parity sector receives contributions solely from the linear order boosts of the Schwarzschild background mass. By contrast, the even-parity sector possesses a dynamical mass aspect that leads to non-trivial supertranslation charges. These supertranslation charges provide an alternative way for fixing the BMS frame at first perturbative order for QNMs.

Indeed, the QNMs of a Schwarzschild black hole can be derived directly within the NU framework \cite{Mongwane:2024vao}. Nevertheless, the linearized Einstein equations expressed in NU gauge are considerably more involved than in the standard Regge-Wheeler-Zerilli formulation \cite{Regge:1957td,Zerilli:1970se}. In particular, the perturbation equations do not reduce to a Schrodinger-type master equation. Instead, they lead to a fourth-order ordinary differential equation \cite{Mongwane:2024vao}, which is less convenient, for instance, for determining the QNM spectrum. By contrast, the explicit coordinate transformations derived in the present work provide an efficient method for extracting the asymptotic NU data associated with generic perturbations. The QNM analysis serves primarily as an illustrative application of the formalism.

The organization of this paper is as follows. In section \ref{Coor-new}, we present the generic algorithm for deriving the coordinate transformations that connect the Boyer–Lindquist (BL) coordinates and the NU coordinates in perturbative expansion. Then, the linear order perturbative metric is obtained in NU gauge. The asymptotic NU data are identified for the perturbations. In section \ref{QNM}, we derive the asymptotic NU data for the QNMs of Schwarzschild black hole. The asymptotic conserved charges are obtained. In particular, the supertranslation charges are applied to fix the BMS frame at first perturbative order. We conclude in the last section. There is also one Appendix to detail the gauge transformation of the perturbative metric.


\section{Algorithm of the coordinate transformation and the linear order perturbative metric in NU gauge}
\label{Coor-new}

The metric of Schwarzschild spacetime in BL coordinates $(t,\rho,\theta,\phi)$ is given by
\be
\td s^2=-\left(1-\frac{2M}{\rho}\right)\td t^2 + \frac{1}{1-\frac{2M}{\rho}} \td \rho^2 + \rho^2 \td \Omega^2.
\ee
For perturbative theory, one considers the Schwarzschild metric as the background solution $\bg_{\mu\nu}$ and the full solution is given by a small parameter $\epsilon$ expansion
\be
g_{\mu\nu}=\bg_{\mu\nu} + \epsilon h_{\mu\nu} + \cO(\epsilon^2).
\ee
Then, the linear order perturbation $h_{\mu\nu}$ satisfies the linearized Einstein equation on the background solution.

We introduce the background NU coordinates $(u_b,r_b,x^A_b)$, which are connected to the BL coordinates as 
\begin{equation}
    t=u_b+\rho_*,\qquad \rho_*=\rho+2M \ln |\frac{\rho-2M}{2M}|,\qquad \rho=r_b.
\end{equation}
The Schwarzschild line-element in the background NU coordinates is given by
\be
\td s^2=-\left(1-\frac{2M}{r_b}\right)\td u_b^2 - 2 \td u_b \td r_b + r^2_b \td \Omega^2.\label{eq:ubrbcordinates}
\ee

Generically, asymptotical flatness of the solution yields that perturbations in Cartesian coordinates $(t,x^i)$ are at order $1/\sqrt{x^ix_i} $ up to gauge freedom. Correspondingly, we assume that the perturbative metric in the background NU coordinates can be expanded at large-$r_b$ as
\begin{align}
&{h}_{r_{b} u_{b}}(r_{b},u_{b},x^{A}_{b})=\sum_{i=1}\frac{1}{r_{b}^i}{h_{i}}_{r_{b} u_{b}}(u_{b},x^{A}_{b}),\nn\\
&{h}_{r_{b} r_{b}}(r_{b},u_{b},x^{A}_{b})=\sum_{i=1}\frac{1}{r_{b}^i}{h_{i}}_{r_{b} r_{b}}(u_{b},x^{A}_{b}) ,\nn\\
&{h}_{u_{b} u_{b}}(r_{b},u_{b},x^{A}_{b})=\sum_{i=1}\frac{1}{r_{b}^i}{h_{i}}_{u_{b} u_{b}}(u_{b},x^{A}_{b}),\label{expansion}\\
&{h}_{r_b A_b}(r_{b},u_{b},x^{A}_{b})={h_{0}}_{r_b A_b}(u_{b},x^{A}_{b})+\sum_{i=1}\frac{1}{r_{b}^i}{h_{i}}_{r_b A_b}(u_{b},x^{A}_{b}),\nn\\
&{h}_{u_b A_b}(r_{b},u_{b},x^{A}_{b})={h_{0}}_{u_b A_b}(u_{b},x^{A}_{b})+\sum_{i=1}\frac{1}{r_{b}^i}{h_{ai}^{(f)}}_{u_b A_b}(u_{b},x^{A}_{b}),\nn\\
&{h}_{A_b B_b}(r_{b},u_{b},x^{A}_{b})= r_{b} {h_{m}}_{A_b B_b}(u_{b},x^{A}_{b})+{h_{0}}_{A_b B_b}(u_{b},x^{A}_{b})+\sum_{i=1}\frac{1}{r_{b}^i}{h_{i}}_{A_b B_b}(u_{b},x^{A}_{b}).\nn
\end{align}

The NU framework consists of both gauge and boundary conditions, which yield the corresponding fall-off conditions of the stress-energy tensor associated to matter fields, see, e.g., discussions in \cite{Pasterski:2021rjz}. Correspondingly, the stress-energy tensor in the background NU gauge should satisfy the following asymptotic behavior
\be\label{stress}
\begin{split}
&T_{r_b r_b}=\cO(r_b^{-4}),\qquad T_{r_b A_b}=\cO(r_b^{-3}),\qquad T_{r_b u_b}=\cO(r_b^{-3}),\\ &T_{u_b u_b}=\cO(r_b^{-2}),\qquad T_{u_b A_b}=\cO(r_b^{-2}),\qquad T_{A_b B_b}=\cO(r_b^{-1}). 
\end{split}
\ee
Otherwise, the metric can not be transformed into NU framework with desired fall-off conditions. Then, the perturbative metric should be consistent with those fall-off conditions via the linearized Einstein equation, which yields that
\be\label{constraints}
\begin{split}
&\p_{u_b} {h_{1}}_{r_b r_b}=0,\qquad \p_{u_b} {h_{0}}_{r_b A_b}=0,\qquad \p_{u_b} [\bg^{AB}{h_{m}}_{A_b B_b}]=0,\\
&{h_{1}}_{u_b r_b}=\frac12 {h_{1}}_{r_b r_b} - \frac12\p_{u_b} {h_{2}}_{r_b r_b}  + \frac14 D^{A_b} D_{A_b} {h_{1}}_{r_b r_b},\\
&{h_{0}}_{u_b A_b}=\frac12 D_{A_b}  {h_{1}}_{u_b r_b} - \frac12 \p_{u_b}  {h_{1}}_{r_b A_b} + \frac12 D^{B_b}  {h_{m}}_{A_b B_b} -\frac12 D_{A_b} (\bg^{{C_b} D_b}{h_{m}}_{C_b D_b}) \\
&\qquad  \qquad\qquad + \frac12 D^2 {h_{0}}_{r_b A_b} - \frac12 D^{B_b} D_{A_b} {h_{0}}_{r_b B_b}  + \frac14D_{A_b} D^2 {h_{1}}_{r_b r_b} +{h_{0}}_{r_b A_b}.
\end{split}
\ee
Those relations are the same as the ones derived in \cite{Mao:2025lwk} based on PM expansion. The reason is that the Schwarzschild spacetime is asymptotically flat and the mass parameter correction is at the subleading order in large-$r_b$ expansion. The above relations are obtained from the leading order linearized equations of motion. Hence, the leading order perturbative metrics on Minkowski and Schwarzschild background share the same constraints from being asymptotic flatness.

We assume that the transformations from the background NU coordinates to the full NU coordinates $(u,r,x^A)$ are also given in the perturbative expansion as
\begin{equation}
\begin{split}
    &u_b=u + \epsilon U_{1}(u,r,x^A)+\epsilon^{2}U_{2}(u,r,x^A)+...,\\
    &r_b=r + \epsilon R_{1}(u,r,x^A)+ \epsilon^{2}R_{2}(u,r,x^A) +...,\\
    &x^{A}_b=x^{A} + \epsilon X^{A}_{1}(u,r,x^A)+ \epsilon^{2}X^{A}_{2}(u,r,x^A) +... .
    \label{transformation-new}
\end{split}
\end{equation}
We apply the following strategy to construct the perturbative diffeomorphism in $\epsilon$ expansion. Starting from the metric in background NU coordinates, the transformed metric is given by
\be\label{transf}
g_{\mu\nu}= g^{(b)}_{\alpha\beta} \frac{\p x_b^\alpha}{\p x^\mu}\frac{\p x_b^\beta}{\p x^\nu}.
\ee
The NU gauge conditions then yield the following transformation laws
\be
g^{(b)}_{\alpha\beta} \frac{\p x_b^\alpha}{\p r}\frac{\p x_b^\beta}{\p r}=0,\qquad g^{(b)}_{\alpha\beta} \frac{\p x_b^\alpha}{\p u}\frac{\p x_b^\beta}{\p r}=-1,\qquad g^{(b)}_{\alpha\beta} \frac{\p x_b^\alpha}{\p x^A}\frac{\p x_b^\beta}{\p r}=0.
\ee
Inserting the relations of the two coordinate systems in small parameter expansion \eqref{transformation-new}, the gauge conditions are reduced to the following equations at the linear order
\be\label{radial}
\partial_r U_1=\frac12 h_{r_b r_b},\, \p_r R_1=h_{u_b r_b} + \frac12 f(r) h_{r_b r_b} - \p_u U_1,\, \p_r X_1^A = \frac{1}{r^2} D^A U_1 - \frac{1}{r^2} {h_{r_b}}^{A_b} ,
\ee
where $f(r)=-1+\frac{2M}{r}$ and $\bar\gamma^{AB}$ is the inverse metric of the celestial sphere in NU coordinates. $D_{A}$ is the covariant derivative associated with the metric $\bar\gamma_{AB}$. Since the perturbative metric $h_{\mu\nu}$ satisfies asymptotically flat conditions, the celestial sphere metric has the same form for background NU and full NU coordinates. The difference between $\bar\gamma_{AB}$ and $\bar\gamma_{A_b B_b}$ is at order $\epsilon$. We are now working at a fixed order in $\epsilon$ expansion. It is not necessary to distinguish the index between $x_b^A$ and $x^A$. From now on the capital indices are raised and lowered by the celestial metric.

The equations in \eqref{radial} are very similar to the case of Minkowski background in \cite{Mao:2025lwk}, except one term in $R_1$ containing the Schwarzschild mass parameter. Nevertheless, the transformed metric components are more distinct as the Schwarzschild mass parameter that arises at the zeroth order will propagate to several places. With respect to the asymptotic expansions \eqref{expansion}, the equations in \eqref{radial} can be solved as
\begin{align}
        U_1&=\frac{1}{2}{h_{1}}_{r_b r_b} \log r +  U_{10}(u,x^A) - \frac{{h_{2}}_{r_b r_b} }{2r}  - \frac{{h_{3}}_{r_b r_b} }{4r^2} + \cO(\frac{1}{r^3}), \label{UG}\\
        X^{A}_1&=X_{10}^A(u,x^A) - \frac{D^A {h_{1}}_{r_b r_b} \log r}{2r} + \frac{2{{h_{0}}_{r_b}}^{A}  - D^A {h_{1}}_{r_b r_b} - 2 D^A U_{10}}{2r}\nn \\
&\qquad +\frac{2 {{h_{1}}_{r_b}}^{A}   +D^A {h_{2}}_{r_b r_b} }{4r^2}  +\frac{4 {{h_{2}}_{r_b}}^{A}  +D^A {h_{3}}_{r_b r_b} }{12r^3} + \cO(\frac{1}{r^4}), \label{XG}\\
        R_1&=-\frac12  \p_u {h_{1}}_{r_b r_b} r \log r + \frac12 \p_u {h_{1}}_{r_b r_b} r- \p_u U_{10} r - \left[ \frac12 {h_{1}}_{r_b r_b} -{h_{1}}_{u_b r_b} - \frac12 \p_u {h_{2}}_{r_b r_b} \right] \log r \nn\\
&\qquad + R_{10}(u,x^A) + \frac{2{h_{2}}_{r_b r_b} - 4{h_{2}}_{u_b r_b} - \p_u {h_{3}}_{r_b r_b} - 4 M {h_{1}}_{r_b r_b} }{4r} +\cO(\frac{1}{r^2}), \label{RG}
\end{align}
where $U_{10}$, $R_{10}$, and $X_{10}^A$ are integration constants with respect to the radial integral. Above, we only compute the terms at the orders that are relevant to asymptotic NU data. The fall-off conditions at large-$r$ of the NU framework are \blue{\cite{Newman:1962cia,Barnich:2011ty}}
\be\label{fall-off}
g_{uA}=o(r),\,\,g_{uu}=-1 + o(1),\,\, g_{AB}=r^2  \bar\gamma_{A B}  + o(r^2),\,\,
|g_{AB}|=r^4|\bg_{AB}|+o(r^3) .
\ee
Those conditions determine the above four integration constants as
\begin{equation}\label{residual}
    \begin{split}
        U_{10}&=T(x^{A})+\frac{u}{2} D_{A}y^{A},\\
        X^{A}_{10}&=y^{A}(x^{C}),\\
        R_{10}&=-\frac14 \bar{\gamma}^{AB}{h_{m}}_{A_{b} B_{b}} + \frac14 D^A D_A {h_{1}}_{r_b r_b} - \frac12 D^A {h_{0}}_{r_b A}  + \frac12 D^2 U_{10},
    \end{split}
\end{equation}
where $T(x^A)$ is the linear order supertranslation and $y^{A}$ satisfy equations
\be
D_{B}y_{A}+D_{A}y_{B}=\bar\gamma_{AB}D_{C}y^{C},
\ee
which are noting but the conformal Killing equations on the celestial sphere. Since the Schwarzschild spacetime is spherically symmetric, it is convenient to split $y^A$ into two parts, namely the linear order rotations with $D_A y^A=0$ and the linear order boosts with $D_A y^A\neq0$. We will show that the linear order rotations will not arise in the final metric in NU gauge.

In \cite{Spiers:2026yqx}, the similar type of transformation was obtained for perturbations in Kerr spacetime. Our results are consistent with theirs. Nevertheless, the aim of that work is to fix the BMS frame for Kerr perturbations and the perturbative metric in BS gauge in terms of the original metric components, as far as we understood, was not explicitly presented.

With the obtained change of coordinates, the components of the perturbative metric in NU gauge at order $\epsilon$ are obtained as
\begin{align}
&{h}_{uu}= \frac{1}{r}\left[ {h_{1}}_{u_b u_b}  + 2 \p_u {h_{2}}_{u_b r_b}  + \frac12 \p_u^2 {h_{3}}_{r_b r_b} + 3 M D_A y^A\right] + \cO(\frac{1}{r^2}), \label{eq:huu}\\
&{h}_{AB}=   \tilde{{C}}_{AB} r \log r +  {C}_{AB} r + \cO(1),  \label{eq:hAB}\\
&\qquad \tilde{{C_1}}_{AB} = -D_A D_B {h_{1}}_{r_b r_b} + \frac12 \bar\gamma_{AB} D^2 {h_{1}}_{r_b r_b} , \\
&\qquad {C}_{AB}= {h_{m}}_{A  B } + 2 D_{(A} {h_{0}}_{B) r_b} + \tilde{{C}}_{AB} - 2 D_A D_B T + \bar\gamma_{AB}D^2 T \nn\\
&\qquad \qquad \qquad \qquad \qquad - \frac12 \bar\gamma_{AB} \bar\gamma^{CD}{h_{m}}_{C  D } - \bar \gamma_{AB} D^C {h_{0}}_{r_b C} , \\
&{h}_{uA}= \bigg(\frac12 D^B \tilde{{C}}_{AB} \log r + \frac12 D^B {{C}}_{AB} - \frac34 D^B  \tilde{{C}}_{AB} \bigg) + \frac{1}{r} \bigg[ {h_{1}}_{u_b A}  + D_A {h_{2}}_{u_b r_b} \nn\\
& + \frac13 \p_u {h_{2}}_{r_b A} + \frac13 \p_u D_A {h_{3}}_{r_b r_b} + M(1+\log r) D_A {h_{1}}_{r_br_b} + 2 M D_A U_{10}\bigg]  + \cO(\frac{1}{r^2}).\label{eq:huA}
\end{align}

To conclude this section, we briefly comment on the order $\epsilon$ metric in NU gauge. There are several logarithmic terms, which are only relevant to ${h_{1}}_{r_b r_b}$. This fully aligns with the smoothness conditions discussed in \cite{Satishchandran:2019pyc}. The extra term $- \frac34 D^B  \tilde{C}_{AB}$ in $g_{uA}$ component at $\cO(r^0)$ is precisely from the logarithmic terms in $g_{AB}$. There is also one logarithmic term at $\cO(r^{-1})$ term, which arises from the order $\cO(r^0)$ terms in $h_{AB}$. That term will not spoil the series expansions at large-$r$ \cite{Barnich:2010eb}, see, also discussions for linearized theory in \cite{Conde:2016rom}.

The Schwarzschild mass contributes only to the Coulombic sector of the gravitational field and does not affect the leading asymptotic shear. The leading shear is determined solely by the radiative data, which somehow reflects the fact that the propagating gravitational degrees of freedom at null infinity are universal across asymptotically flat spacetimes. The news tensor
\be
{N}_{AB}=\p_u {C}_{AB}=\p_{u_b} {h_{m}}_{A  B }  - \frac12 \bar\gamma_{AB} \bar\gamma^{CD} \p_{u_b} {h_{m}}_{C  D } ,
\ee
is completely fixed by the traceless part of the linear transverse metric in the background NU coordinates.

For the mass aspect,
\be\label{mass}
m= {h_{1}}_{u_b u_b}  + 2 \p_u {h_{2}}_{u_b r_b}  + \frac12 \p_u^2 {h_{3}}_{r_b r_b} + 3 M D_A y^A,
\ee
it receives contributions from the Schwarzschild mass through the linear order boosts $y^A$, which is consistent with the results, e.g., in the Appendix of \cite{Bonga:2018gzr}. For the angular momentum aspect,  
\begin{multline}\label{AM}
{N}_A= {h_{1}}_{u_b A}  + D_A {h_{2}}_{u_b r_b}  + \frac13 \p_u {h_{2}}_{r_b A} + \frac13 \p_u D_A {h_{3}}_{r_b r_b} \\
+ M D_A {h_{1}}_{r_br_b} + 2 M D_A T + u M D_A D_B y^B,
\end{multline}
there is an interesting extra piece arising from the ${h_{1}}_{r_br_b}$ term, which indicates that the non-smoothness configuration may lead to physical effect through the angular momentum distribution. Nevertheless, all the corrections from the Schwarzschild mass are given as total derivative terms which do not affect the definition of global energy and angular momentum of the perturbations.


\section{QNMs in the NU gauge}
\label{QNM}

Quasinormal modes are resonant solutions of the black hole perturbation equations. They are important because their frequencies and damping rates are fixed by the background geometry and therefore give a simple characterization of the black hole response \cite{Vishveshwara:1970zz,Press:1971wr,Chandrasekhar:1975zza,Leaver:1985ax}. QNMs have acquired renewed observational importance with the development of gravitational-wave astronomy. The ringdown signal of a binary black hole merger provides a direct probe of the remnant black hole. If several QNMs can be resolved, their frequencies and damping times can be compared with the spectrum predicted for the black hole, which is the basic idea of black hole spectroscopy \cite{Dreyer:2003bv,Berti:2005ys,Isi:2019aib,Giesler:2019uxc}. For instance, the post-merger ringdown signal was used to test the Kerr QNM spectrum \cite{LIGOScientific:2025wao}.

We work in the background NU coordinates $(u_b,r_b,x^A_b)$ introduced in Eq.~\eqref{eq:ubrbcordinates}, where the Schwarzschild metric can be organized into a warped-product form as
\be
\td s^2 = g_{ij}\,\td y^i\td y^j + r_b^2\bar\gamma_{AB}\,\td x_b^A\td x_b^B, \qquad y^i=(u_b,r_b),
\ee
with
\be
g_{ij}=
\begin{pmatrix}
-f & -1\\
-1 & 0
\end{pmatrix},
\qquad
f(r_b)=1-\frac{2M}{r_b}.
\ee
Indices $i,j$ refer to the two-dimensional orbit space $\mathcal M^2$, while $A,B$ refer to the unit two-sphere. We denote the corresponding covariant derivatives by $\nabla_i$ and $D_A$, respectively.

Since the Schwarzschild spacetime is spherically symmetric. It is very convenient to apply spherical harmonic expansion. For a fixed $(\ell,m)$ mode, let $Y_{\ell m}$ be the real scalar spherical harmonic on the unit two-sphere, and let $Y:=Y_{\ell m}$ denote arbitrary mode. We define the derived vector and tensor harmonics by
\be
Y_A=D_A Y, \qquad X_A=-\epsilon_A{}^B D_B Y,
\ee
and
\be
Y_{AB}=\left(D_A D_B+\frac12\ell(\ell+1)\bar\gamma_{AB} \right)Y, \qquad X_{AB}=D_{(A}X_{B)}.
\ee
Here $\epsilon_{AB}$ is the Levi--Civita tensor on the unit two-sphere, and $\epsilon_A{}^B=\bar\gamma^{BC}\epsilon_{AC}$. All angular indices are raised and lowered with $\bar\gamma_{AB}$.

For the $\ell\geq2$ case, the metric perturbation separates into even- and odd-parity sectors. The two parity sectors can be written in block form as \cite{Martel:2005ir}
\be
h_{\mu\nu}^{\mathrm e}=
\begin{pmatrix}
H_{ij}Y & j_iY_B \\[1mm]
& r_b^2\left(K\bar\gamma_{AB}Y+GY_{AB}\right)
\end{pmatrix},
\qquad
h_{\mu\nu}^{\mathrm o}=
\begin{pmatrix}
0  & h_iX_B \\[1mm]
& h_2X_{AB}
\end{pmatrix}.
\label{eq:paritydecomposition}
\ee
At linear order, the two perturbations \eqref{eq:paritydecomposition} decouple and therefore define two independent parity sectors.

The low multipoles $\ell=0,1$ have a very different character. In the vacuum configuration, the $\ell=0$ even-parity perturbation corresponds to a variation of the Schwarzschild mass, while the $\ell=1$ odd-parity perturbation corresponds to an infinitesimal angular-momentum perturbation, i.e. the linearized Kerr solution. The $\ell=1$ even-parity sector is pure gauge \cite{Martel:2005ir}. These modes are non-radiative. Since we are interested in gravitational QNMs, we restrict from now on to $\ell\geq2$.

The Regge-Wheeler gauge \cite{Regge:1957td,Martel:2005ir} is widely applied to eliminate the gauge freedom of the perturbations, which is specified by $j_i=0=G$ and $h_2=0$ for even and odd-parities, respectively. In each fixed $(\ell,m)$ mode this completely fixes the gauge freedom and considerably simplifies the linearized field equations. Nevertheless, it is important to organize the perturbative equations in terms of gauge-invariant quantities. Define
\be
\lambda_{\rm o}:=\ell(\ell+1),\qquad \lambda_{\rm e}:=\frac12(\ell-1)(\ell+2),\qquad \Lambda:=2\lambda_{\rm e}+\frac{6M}{r_b}.
\label{eq:angularparameters}
\ee
The odd-parity Cunningham-Price-Moncrief (CPM) variable can be introduced \cite{Cunningham:1978zfa,Martel:2005ir}
\be
\Psi_{\rm o}:=\frac{r_b^3}{\lambda_{\rm e}} \epsilon^{ij}\nabla_i\left(\frac{\widetilde h_j}{r_b^2}\right),
\label{eq:CPMsclar}
\ee
where $\epsilon^{ij}$ is the volume element associated with $g_{ij}$. We choose the orientation $\epsilon_{u_br_b}=+1$. The opposite orientation changes the sign of all these Levi–Civita components and correspondingly changes the overall sign convention of the odd-parity master variable. For the even-parity case, the Zerilli-Moncrief (ZM) variable is defined as \cite{Martel:2005ir,Moncrief:1974am}
\be
\Psi_{\rm e}:=\frac{2r_b}{\lambda_{\rm o}} \left[\widetilde K+\frac{2}{\Lambda} \left(r^ir^j\widetilde h_{ij} -r_b r^i\nabla_i\widetilde K\right)\right].
\label{eq:ZMscalar}
\ee
Variables with an overhead tilde are defined as \cite{Martel:2005ir}
\begin{align}
&\widetilde h_{ij}:=H_{ij}-2\nabla_{(i}\varepsilon_{j)},\qquad
\widetilde K:=K+\frac12\ell(\ell+1)G-\frac{2}{r_b}r^i\varepsilon_i,\\
&\widetilde h_i:=h_i-\frac12\nabla_i h_2+\frac{1}{r_b}r_i h_2,
\label{eq:gaugeinvariants}
\end{align}
where $r_i=\nabla_i r_b$ and $r^i=g^{ij}r_j$.

The two master variables satisfy the Regge-Wheeler-Zerilli equations for the source-free case \cite{Regge:1957td,Zerilli:1970se},
\be
\left[-\partial_t^2+\partial_{r_*}^2-V_p(r_b)\right]\Psi_p=0, \qquad p\in\{\mathrm{o},\mathrm{e}\}.
\label{eq:RWZmaster}
\ee
The corresponding potentials are
\begin{align}
V_{\mathrm{o}}(r_b) &=V_{\mathrm{RW}}(r_b) =f\left[\frac{\lambda_{\mathrm{o}}}{r_b^2}-\frac{6M}{r_b^3}\right], \label{eq:RWpotential}\\
V_{\mathrm{e}}(r_b) &=V_{\mathrm{Z}}(r_b) =f\frac{ 2\lambda_{\mathrm{e}}^2(\lambda_{\mathrm{e}}+1)r_b^3 +6\lambda_{\mathrm{e}}^2Mr_b^2 +18\lambda_{\mathrm{e}}M^2r_b +18M^3 }{ r_b^3(\lambda_{\mathrm{e}}r_b+3M)^2 }. \label{eq:Zpotential}
\end{align}

The metric components in the Regge-Wheeler can be reconstructed from the master variables \cite{Sarbach:2001qq,Martel:2003jj,Buchman:2007pj}. Specifically, the odd-parity sector is reconstructed as
\be
h_i^{\rm RW} = \frac12\epsilon_i{}^j\nabla_j(r_b\Psi_{\rm o}).
\label{eq:oddreconstruction}
\ee
For the even-parity sector, one has
\begin{align}
K^{\rm RW} &=\frac{1}{\Lambda}\left[ r^i\nabla_i(\Lambda\Psi_{\rm e}) +\frac{\lambda_{\rm o}\Lambda}{2r_b}\Psi_{\rm e} \right], \label{eq:evenreconstructionK} \\
H_{ij}^{\rm RW} &=\frac{1}{\Lambda} \left( \nabla_i\nabla_j-\frac12 g_{ij}\nabla^k\nabla_k \right) (r_b\Lambda\Psi_{\rm e}) . \label{eq:evenreconstructionH}
\end{align}
In the $(u_b,r_b)$ coordinates the reconstruction formulas are
\begin{align}
h_{r_b}^{\rm RW} &=\frac12\left(\Psi_{\rm o}+r_b\partial_{r_b}\Psi_{\rm o}\right), \qquad h_{u_b}^{\rm RW} =-\frac{r_b}{2}\partial_{u_b}\Psi_{\rm o}+f h_{r_b}^{\rm RW}, \label{eq:oddreconstructionEF} \\[1mm]
K^{\rm RW} &=-\partial_{u_b}\Psi_{\rm e} +f\partial_{r_b}\Psi_{\rm e} +\left(\frac{\lambda_{\rm e}+1}{r_b} -\frac{6Mf}{r_b^2\Lambda}\right)\Psi_{\rm e}, \label{eq:evenreconstructionKEF} \\[1mm]
H_{r_br_b}^{\rm RW} &=r_b\partial_{r_b}^2\Psi_{\rm e} +\frac{4\lambda_{\rm e}}{\Lambda}\partial_{r_b}\Psi_{\rm e}, \qquad H_{u_br_b}^{\rm RW} =\frac f2 H_{r_br_b}^{\rm RW}, \label{eq:evenreconstructionradial} \\[1mm]
H_{u_bu_b}^{\rm RW} &=r_b\partial_{u_b}^2\Psi_{\rm e} -fr_b\partial_{u_b}\partial_{r_b}\Psi_{\rm e} +\left(\frac{M}{r_b}-\frac{2f\lambda_{\rm e}}{\Lambda}\right) \partial_{u_b}\Psi_{\rm e} +\frac{f^2}{2}H_{r_br_b}^{\rm RW}. \label{eq:evenreconstructionuu}
\end{align}

One can further separate the time dependence of the master field as
\be
\Psi_p(t,r_b)=e^{-i\omega t}\psi_p(r_b), \qquad p\in\{\mathrm{o},\mathrm{e}\}.
\label{eq:QNMseparation}
\ee
The complex form in \eqref{eq:QNMseparation} is only a convenient mode representation. After substituting $\Psi_p$ into the reconstruction formulas above, the physical metric perturbation is obtained by taking $\operatorname{Re}h_{\mu\nu}^{\rm RW}$. The QNMs are defined by purely ingoing behavior at the black hole horizon and purely outgoing behavior at infinity,
\begin{equation}
\psi \sim e^{-i\omega \rho_*}, \qquad \rho_*\to-\infty, \qquad \psi \sim e^{+i\omega \rho_*}, \qquad \rho_*\to+\infty .
\label{eq:QNM-boundary-short}
\end{equation}
The odd- and even-parity sectors share the same gravitational QNM spectrum \cite{Chandrasekhar:1975zza}, so we suppress the parity label on $\omega$. The frequency $\omega$ is not an arbitrary input parameter. They are the discrete complex eigenvalues (resonances) of the linearized perturbation operator on the Schwarzschild background. They depend only on the parameters of the background spacetime and the chosen harmonic indices $(\ell,m)$, and are independent of the particular perturbation used to excite the modes. They are fixed from the desired boundary conditions and are known as the QNM spectrum. Accurate numerical values of the spectrum may be obtained, for example, by the continued-fraction method \cite{Leaver:1985ax,Nollert:1993zz}. The master equation \eqref{eq:RWZmaster} then reduces to the one-dimensional Schrodinger-type radial equation
\be
\frac{\mathrm d^2\psi_p}{\mathrm d \rho_*^2} +\left[\omega^2-V_p(r_b)\right]\psi_p=0 .
\label{eq:QNMradial}
\ee

In the present work, we mainly focus on the asymptotic NU data for QNMs, which are controlled by asymptotic data of the QNM solutions as demonstrated in previous section. We therefore introduce series expansions at infinity as
\be
\psi_p(r_b) = e^{i\omega \rho_*}\sum_{n=0}^{\infty}a_n^p r_b^{-n}, \qquad a_0^p=1 .
\label{eq:QNMseries}
\ee
The choice $a_0^p=1$ fixes the overall normalization of the outgoing solution.

For the large-$r_b$ analysis, it is also convenient to expand the potentials in \eqref{eq:RWpotential} and \eqref{eq:Zpotential} as
\be
V_p(r_b)=\sum_{j=2}^{\infty}v_j^p r_b^{-j}.
\label{eq:Vseries}
\ee
For the odd-parity Regge--Wheeler potential, the expansion terminates,
\be
v_2^{\rm o}=\lambda_{\rm o},\qquad v_3^{\rm o}=-2M(\lambda_{\rm o}+3),\qquad v_4^{\rm o}=12M^2,\qquad v_j^{\rm o}=0\quad (j\geq5).
\label{eq:Vcoeffodd}
\ee
For the even-parity Zerilli potential, the first four order coefficients are
\begin{align}
v_2^{\rm e} &=2(\lambda_{\rm e}+1),\qquad v_3^{\rm e} =-\frac{2M(2\lambda_{\rm e}^2+5\lambda_{\rm e}+6)} {\lambda_{\rm e}}, \\
v_4^{\rm e} &=\frac{6M^2(2\lambda_{\rm e}^2+10\lambda_{\rm e}+9)} {\lambda_{\rm e}^2}, \qquad v_5^{\rm e} =-\frac{36M^3(\lambda_{\rm e}+2)(2\lambda_{\rm e}+3)} {\lambda_{\rm e}^3}. \label{eq:Vcoeffeven}
\end{align}
More generally, for $j\geq5$,
\be
v_j^{\rm e} = \frac{2}{9} \left(-\frac{3M}{\lambda_{\rm e}}\right)^{j-2} \left[ 4\lambda_{\rm e}^2(j-2) +6\lambda_{\rm e}(2j-3) +9(j-1) \right].
\label{eq:Vcoeffevengeneral}
\ee
Substituting the large-$r_b$ expansions \eqref{eq:QNMseries} and \eqref{eq:Vseries} into the radial equation \eqref{eq:QNMradial}, and matching equal powers of $r_b^{-1}$, yield the following recurrence relations
\begin{align}
2i\omega k\,a_k^p ={}& (k-1)(k+4iM\omega)a_{k-1}^p -2M(k-2)(2k-1)a_{k-2}^p \nonumber\\
&\qquad + 4M^2(k-3)(k-1)a_{k-3}^p -\sum_{j=2}^{k+1}v_j^p a_{k+1-j}^p, \qquad k\geq1 . \label{eq:QNMrecurrence}
\end{align}
Here $a_n^p=0$ for $n<0$. This relation determines the large-$r_b$ expansion recursively for each fixed nonzero $\omega$ of both parity sectors. We will reconstruct the perturbative metric for different sectors and obtain the corresponding asymptotic NU data in the following subsections.

\subsection{Parity odd sector}

We first consider the simpler odd-parity reconstruction. Substituting the series expansions into the odd-parity reconstruction formulas \eqref{eq:oddreconstructionEF} gives
\begin{align}
h_{r_b}^{\rm RW} &= \frac{e^{-i\omega u_b}}{2} \sum_{n=0}^{\infty}(1-n)a_n^{\rm o}r_b^{-n}, \label{eq:oddseriesr} \\
h_{u_b}^{\rm RW} &= \frac{e^{-i\omega u_b}}{2} \sum_{n=0}^{\infty} \left[ i\omega a_n^{\rm o} +(2-n)a_{n-1}^{\rm o} -2M(3-n)a_{n-2}^{\rm o} \right]r_b^{1-n}. \label{eq:oddseriesu}
\end{align}
The second equation contains a leading term of order $\mathcal{O}(r_b)$, namely $h_{u_b}^{\rm RW}\sim r_b e^{iu_b\omega}$. Thus $h_{u_bA}^{\rm RW}$ does not satisfy the large-$r_b$ behavior assumed in the expansion~\eqref{expansion}. Nevertheless, the over-leading term can be removed by a gauge transformation before performing the asymptotic expansion. Using the odd-parity gauge transformation \eqref{eq:oddgaugetransformation} in Appendix \ref{gauge} and choosing
\be
\xi^{\rm o}=-\frac{r_b}{2}\Psi_{\rm o},
\label{eq:oddNUgauge}
\ee
one obtains
\be
h_i = \frac12\epsilon_i{}^j\nabla_j(r_b\Psi_{\rm o}) +\frac{r_b}{2}\nabla_i\Psi_{\rm o} -\frac12 r_i\Psi_{\rm o}, \qquad h_2=r_b\Psi_{\rm o},
\label{eq:oddNUreconstruction}
\ee
from the reconstruction formula \eqref{eq:oddreconstruction}.
 
In the $(u_b,r_b)$ coordinates these expressions are reduced to
\be
h_{r_b}=r_b\partial_{r_b}\Psi_{\rm o}, \qquad h_{u_b}=\frac{f}{2}\partial_{r_b}(r_b\Psi_{\rm o}), \qquad h_2=r_b\Psi_{\rm o}.
\label{eq:oddNUcomponents}
\ee
To compare with the expansion in \eqref{expansion}, we define
\be
\omega=\omega_{\rm R}+i\omega_{\rm I},\quad \Delta_\omega=\omega_{\rm R}^2+\omega_{\rm I}^2, \quad c_\omega(u_b)=\cos(\omega_{\rm R}u_b),\quad s_\omega(u_b)=\sin(\omega_{\rm R}u_b),
\ee
where $\omega_{\rm I}<0$ for a damped QNM. Inserting the coefficients from the recursive relation in \eqref{eq:QNMrecurrence}, we obtain non-vanishing odd-parity metric components in the background NU coordinates as
\begin{align}
h_{u_bA}^{\rm o} &=e^{\omega_{\rm I}u_b} \left[\frac12 c_\omega(u_b)-\frac{M}{r_b}c_\omega(u_b) +\mathcal O(r_b^{-2})\right]X_A, \label{eq:oddmetricuA}\\
h_{r_bA}^{\rm o} &=-\frac{\lambda_{\rm o}e^{\omega_{\rm I}u_b}}{2\Delta_\omega r_b} \left[\omega_{\rm I}c_\omega(u_b)+\omega_{\rm R}s_\omega(u_b)\right]X_A +\frac{h_{2r_bA}^{\rm o}}{r_b^2} +\mathcal O(r_b^{-3}), \label{eq:oddmetricrA}\\
h_{AB}^{\rm o} &=e^{\omega_{\rm I}u_b} \left[r_b c_\omega(u_b) +\frac{\lambda_{\rm o}}{2\Delta_\omega} \left(\omega_{\rm I}c_\omega(u_b)+\omega_{\rm R}s_\omega(u_b)\right) +\mathcal O(r_b^{-1})\right]X_{AB}, \label{eq:oddmetricAB}
\end{align}
where
\begin{align}
h_{2r_bA}^{\rm o}&=\frac{e^{\omega_{\rm I}u_b}}{2\Delta_\omega^2}\Big[ \big\{\lambda_{\rm o}\lambda_{\rm e} (\omega_{\rm R}^2-\omega_{\rm I}^2) +6M\omega_{\rm I}\Delta_\omega\big\}c_\omega(u_b) \nonumber\\
&\hspace{30mm} +2\omega_{\rm R} \big(3M\Delta_\omega -\lambda_{\rm o}\lambda_{\rm e}\omega_{\rm I}\big) s_\omega(u_b) \Big]X_A.
\end{align}
Now, one can extract the asymptotic NU data from \eqref{eq:huu}-\eqref{eq:huA} as
\be\label{odd}
\begin{split}
&C_{AB}=e^{\omega_{\rm I}u} c_\omega(u) X_{AB}  - 2 D_A D_B T + \bar\gamma_{AB}D^2 T , \qquad m=3 M D_A y^A,\\
& N_A=-\frac{\lambda_{\rm o}\lambda_{\rm e}e^{\omega_{\rm I}u} }{6 \Delta_\omega} \left[\omega_{\rm I}c_\omega(u)+\omega_{\rm R}s_\omega(u)\right] X_A + 2 M D_A T + u M D_A D_B y^B.
\end{split}
\ee
In particular, the mass aspect is nothing but the linear order boosts of the Schwarzschild mass. Since $y^A$ is a conformal Killing vector of the celestial sphere, one can expand $D_Ay^A$ by $\ell=1$ spherical harmonics, namely $D_Ay^A= a_{-1} Y_{1,-1} +a_{0} Y_{1,0} + a_{1} Y_{1,1}$, which yields a vanishing total energy after a sphere integral from the orthogonality of harmonics. There is no dynamical mass aspect associated to the odd-parity sector, hence, the parity odd sector corresponds to the magnetic sector. While the angular momentum aspect has dynamical piece. But the sphere integral of the angular momentum aspect yields a zero total angular momentum at the linear order. The radiated energy and radiated angular momentum start at least at the second perturbative order.

As a final remark, one can verify that the above asymptotic NU data satisfy the desired relations from the linearized equations of motion in NU gauge,
\be\label{NU}
\begin{split}
&\p_u m=\frac12\p_u (D^A D^B C_{AB})+ \frac12 \p_u (\gamma^{AB} C_{AB})=0,\\
&\p_u N_A=\frac13 D_A m + \frac16 D^B D_A D^C C_{BC} - \frac16 D^2 D^B C_{AB}.
\end{split}
\ee
Actually, the asymptotic NU data \eqref{odd} can be alternatively obtained from physical requirements and the above evolution relations. Schwarzschild spacetime is spherically symmetric and $\frac{\p }{\p u}$ is a Killing vector. So the asymptotic magnetic sector shear tensor should be organized in the form of $e^{\omega_{\rm I}u} c_\omega(u) X_{AB}$. The only freedom is the normalization at the infinity. Once the asymptotic shear tensor is given, the mass aspect and angular momentum aspect can be fixed by the evolution relations in \eqref{NU} up to integration constants. Since QNMs are resonant solutions, the total energy and angular momentum are zero. Hence, the integration constants are purely from the gauge freedom which is given in \eqref{mass} and \eqref{AM}. Nevertheless, our computations present the precise formulas and confirm the physical understanding of QNMs in the NU gauge. Note also that the asymptotic NU data are obtained for generic perturbations of Schwarzschild in above section. The QNM computations just provide a direct application. For the parity even sector as will be detailed in the next subsection, the mass aspect includes dynamical terms. Consequently, the QNMs have non-trivial supertranslation charges, which offers a simple setup to understand the physical effect of supertranslation charges.

\subsection{Parity even sector}

We now turn to metric reconstruction for the even-parity sector. Substituting the master-field expansion \eqref{eq:QNMseries} into \eqref{eq:evenreconstructionK} and \eqref{eq:evenreconstructionH} yields the corresponding large-$r_b$ expansions,
\begin{align}
K^{\rm RW} =e^{-i\omega u_b}\sum_{n=0}^{\infty} \Bigg[ &i\omega a_n^{\rm e} +(\lambda_{\rm e}+2-n)a_{n-1}^{\rm e} +2M(n-2)a_{n-2}^{\rm e} \nonumber\\
&+\sum_{m=1}^{n-1}q_m a_{n-m-1}^{\rm e} -2M\sum_{m=1}^{n-2}q_m a_{n-m-2}^{\rm e} \Bigg]r_b^{-n}. \label{eq:evenRWKseries}
\end{align}
and
\begin{align}
H_{r_br_b}^{\rm RW} &=e^{-i\omega u_b}\sum_{n=2}^{\infty} \left[ n(n-1)a_n^{\rm e} -2\sum_{m=1}^{n-1}(n-m)q_m a_{n-m}^{\rm e} \right]r_b^{-n-1}, \label{eq:evenRWrrseries}\\
H_{u_br_b}^{\rm RW} &=\frac{f}{2}H_{r_br_b}^{\rm RW}. \label{eq:evenRWurseries}\\
H_{u_bu_b}^{\rm RW} &={}e^{-i\omega u_b} \Bigg[ -\omega^2\sum_{n=0}^{\infty}a_n^{\rm e}r_b^{1-n} -i\omega f\sum_{n=0}^{\infty}n a_n^{\rm e}r_b^{-n} -iM\omega\sum_{n=0}^{\infty}a_n^{\rm e}r_b^{-n-1} \nonumber\\
&\hspace{19mm} +i\omega f\sum_{m,n=0}^{\infty}q_m a_n^{\rm e}r_b^{-m-n} \Bigg] +\frac{f^2}{2}H_{r_br_b}^{\rm RW}, \label{eq:evenRWuuseries}
\end{align}
where we define
\be
\frac{1}{\Lambda} = \frac{1}{2\lambda_{\rm e}} \sum_{n=0}^{\infty}q_n r_b^{-n}, \qquad q_n=\left(-\frac{3M}{\lambda_{\rm e}}\right)^n ,
\ee
to keep the resulting coefficients in compact forms. However, those results yield that
\be
h_{u_bu_b}^{\rm RW}=\mathcal O(r_b), \qquad h_{AB}^{\rm RW}=\mathcal O(r_b^2),
\ee
in the background NU coordinates, which breaks the fall-off conditions in \eqref{expansion}.

Nevertheless, one can apply a gauge transformation generated by
\begin{align}
\xi^{\rm e} &= \frac{H_{u_bu_b}^{\rm RW}+\frac{r_b}{2}(\partial_{r_b}f)K^{\rm RW}}{2\omega^2-\frac{\lambda_{\rm o}}{2r_b}\partial_{r_b}f}, \label{eq:evenNUxie}\\
\xi_{u_b}&=i\omega\xi^{\rm e}, \qquad \xi_{r_b}=\frac1f\left[ \frac{r_b}{2}K^{\rm RW}+i\omega\xi^{\rm e}+\frac{\lambda_{\rm o}}{2r_b}\xi^{\rm e} \right]. \label{eq:evenNUxi}
\end{align}
The asymptotic behavior of the transformed fields are
\begin{align}
&j_{u_b}=0,\qquad j_{r_b}=\mathcal O(r_b^{-1}),\qquad K=0. \\
&H_{u_bu_b}=0,\qquad H_{u_br_b}=\mathcal O(r_b^{-1}),\qquad H_{r_br_b}=\mathcal O(r_b^{-2}).
\end{align}
We again take the real part of the even-parity radial variable. Inserting the coefficients from the recursive relation in \eqref{eq:QNMrecurrence}, we obtain the non-vanishing even-parity metric components as
\begin{align}
h_{u_br_b}^{\rm e} &=\frac{\lambda_{\rm e}(\lambda_{\rm e}+1)e^{\omega_{\rm I}u_b}} {2\Delta_\omega r_b^2} \left[\omega_{\rm I}c_\omega(u_b)+\omega_{\rm R}s_\omega(u_b)\right]Y+\mathcal O(r_b^{-3}), \label{eq:evenmetricur}\\
h_{r_br_b}^{\rm e} &=\mathcal O(r_b^{-4}), \label{eq:evenmetricrr}\\
h_{r_bA}^{\rm e} &=-\frac{\lambda_{\rm e}e^{\omega_{\rm I}u_b}}{\Delta_\omega r_b} \left[\omega_{\rm I}c_\omega(u_b)+\omega_{\rm R}s_\omega(u_b)\right]Y_A +\frac{h_{2r_bA}^{\rm e}}{r_b^2} +\mathcal O(r_b^{-3}), \label{eq:evenmetricrA}\\
h_{AB}^{\rm e} &=e^{\omega_{\rm I}u_b} \left[ r_b c_\omega(u_b) +\frac{\lambda_{\rm e}}{\Delta_\omega} \left(\omega_{\rm I}c_\omega(u_b)+\omega_{\rm R}s_\omega(u_b)\right) +\mathcal O(r_b^{-1}) \right]Y_{AB}. \label{eq:evenmetricAB}
\end{align}
Here
\begin{align}
h_{2r_bA}^{\rm e} &=\frac{\lambda_{\rm e}(\lambda_{\rm e}+1)e^{\omega_{\rm I}u_b}} {2\Delta_\omega^2} \left[ (\omega_{\rm R}^2-\omega_{\rm I}^2)c_\omega(u_b) -2\omega_{\rm R}\omega_{\rm I}s_\omega(u_b) \right]Y_A. \label{eq:evenh2rA}
\end{align}
Now, one can extract the asymptotic NU data from \eqref{eq:huu}-\eqref{eq:huA} as
\be
\begin{split}
&C_{AB}=e^{\omega_{\rm I}u} c_\omega(u) Y_{AB}  - 2 D_A D_B T + \bar\gamma_{AB}D^2 T , \\
&m=\lambda_{\rm e}(\lambda_{\rm e}+1)  c_\omega(u) e^{\omega_{\rm I}u} Y + 3 M D_A y^A,\label{M}\\
& N_A=\frac{\lambda_{\rm e}(\lambda_{\rm e}+1)e^{\omega_{\rm I}u}} {3\Delta_\omega } \left[\omega_{\rm I}c_\omega(u)+\omega_{\rm R}s_\omega(u)\right] Y_A  + 2 M D_A T + u M D_A D_B y^B.
\end{split}
\ee
One can verify that those data satisfy the relations in \eqref{NU} as they should be. Again, the total mass and angular momentum are zero at the linear order, and the radiated energy and radiated angular momentum start at least at the second perturbative order.

\subsection{Fixing the BMS frame at first perturbative order}

Interestingly, the mass aspect of the parity even sector is non-trivial, which contributes to the supertranslation charges defined at the null infinity. Let $S(x^A)$ be the supertranslation parameter,\footnote{Here, the supertranslation is not the arbitrary function $T(x^A)$ on celestial sphere in \eqref{residual} which is at order $\epsilon$. It is the supertranslation of the background Schwarzschild spacetime, e.g., \cite{Hawking:2016sgy}.} the corresponding charge is given by \cite{Barnich:2011mi}
\be
Q_S=\int \td \Omega^2 S m.
\ee
Expanding the supertranslation parameter as 
\be
S=\sum_{\ell,m}S_{\ell,m}Y_{\ell,m},
\ee
and inserting the complete spherical harmonics in the mass aspect in \eqref{M},
\be
Y=\sum_{\ell,m}C_{\ell,m}Y_{\ell,m},
\ee 
yield the $(\ell,m)$ mode supertranslation charge as
\be\label{supertranslation}
Q_{\ell,m}=C_{\ell,m} \lambda_{\rm e}(\lambda_{\rm e}+1)  c_\omega(u) e^{\omega_{\rm I}u},\qquad \ell\geq 2.
\ee

The supertranslation charge enters naturally in the problem of fixing the BMS frame at first perturbative order. In \cite{Spiers:2026yqx}, it was proposed that the linear order supertranslation $T(x^A)$ should be fixed by requiring the even-parity part of the shear $C_{AB}$ to vanish at a chosen reference time. Expanding the supertranslation parameter in harmonics
\be
T=\sum_{\ell,m}T_{\ell,m}Y_{\ell,m},
\ee 
this prescription determines the coefficients $T_{\ell,m}$ in terms of the same perturbation amplitudes $C_{\ell,m}$ according to the explicit formula of the even-parity shear in \eqref{M}. Crucially, the perturbation amplitudes $C_{\ell,m}$ uniquely determine the supertranslation charge modes at a chosen reference time in \eqref{supertranslation}. Consequently,
\begin{equation}
T_{\ell,m}\sim C_{\ell,m}\sim Q_{\ell,m}.
\end{equation}
Remarkably, the first order supertranslation can be alternatively fixed by the supertranslation charge at this order. This provides the final ingredient in fixing the BMS frame formulated entirely in terms of asymptotic charges \cite{Spiers:2026yqx}. We expect that this prescription can be valid for generic perturbation theory applying the NU relations in \eqref{NU}.

\section{Concluding remarks}

In this paper, we derive the explicit coordinate transformations that map generic first order metric perturbations of a Schwarzschild background into the NU framework and extract the corresponding asymptotic NU data. As a direct application, we study the QNMs of a Schwarzschild black hole in detail. The total energy and angular momentum obtained from the asymptotic NU data vanish at the linear order, which is consistent with the resonant nature of QNMs. The mass aspect in the odd-parity sector consists solely of a linear order boost of the Schwarzschild background mass. Whereas the even-parity sector contains a dynamical contribution to the mass aspect. Consequently, the supertranslation charges are non-trivial in this sector.   

The supertranslation charge inherits the same oscillation frequency and damping rate as the corresponding QNM. This observation may be interesting from the perspective of semiclassical gravity, where the partition function of finite temperature spacetimes can be related to the QNM spectrum of the corresponding black hole backgrounds \cite{Denef:2009kn}. The connection between semiclassical quantum gravity and the representation theory of asymptotic symmetry groups is well established in lower-dimensional gravity \cite{Strominger:1997eq,Witten:2007kt,Yin:2007gv,Maloney:2007ud,Giombi:2008vd,Oblak:2015sea,Barnich:2015mui}. Our explicit formulas for the non-trivial supertranslation charges associated with Schwarzschild QNMs may therefore provide a new perspective on the interplay between QNM spectra, asymptotic symmetries, and semiclassical quantum gravity in higher-dimensional spacetimes.


\section*{Acknowledgments}

This work is supported in part by the National Natural Science Foundation of China (NSFC) under Grants No.~12475059 and No.~11935009, and by Tianjin University Self-Innovation Fund Extreme Basic Research Project Grant No.~2025XJ21-0007. K.-Y. Zhang acknowledges financial support from the China Scholarship Council (CSC).

\appendix

\section{The gauge freedom of metric perturbations }
\label{gauge}

For a linear perturbative gauge transformation $h_{\mu\nu}\rightarrow h_{\mu\nu}-\mathcal L_\xi\bar g_{\mu\nu}$, generated by $\xi^\mu$, the gauge parameter can be decomposed for each $(\ell,m)$ mode as
\be
\xi_\mu^{\rm e}\td x^\mu=\xi_iY\,\td y^i+\xi^{\rm e}Y_A\,\td x_b^A,\qquad
\xi_\mu^{\rm o}\td x^\mu=\xi^{\rm o}X_A\,\td x_b^A .
\label{eq:gaugegenerator}
\ee
The corresponding variables are transformed as
\begin{align}
H_{ij}&\rightarrow H_{ij}-2\nabla_{(i}\xi_{j)},&
j_i&\rightarrow j_i-\xi_i-\nabla_i\xi^{\rm e}+\frac{2}{r_b}r_i\xi^{\rm e}, \\
G&\rightarrow G-\frac{2}{r_b^2}\xi^{\rm e},&
K&\rightarrow K+\frac{\ell(\ell+1)}{r_b^2}\xi^{\rm e}-\frac{2}{r_b}r^i\xi_i, \label{eq:evengaugetransformation}\\
h_i&\rightarrow h_i-\nabla_i\xi^{\rm o}+\frac{2}{r_b}r_i\xi^{\rm o},&
h_2&\rightarrow h_2-2\xi^{\rm o}.
\label{eq:oddgaugetransformation}
\end{align}


\providecommand{\href}[2]{#2}\begingroup\raggedright\endgroup

\end{document}